\documentclass[prd,onecolumn,amsmath,amssymb,superscriptaddress,nofootinbib,10pt]{revtex4-2}
\usepackage[T1]{fontenc}
\usepackage{lmodern}
\usepackage{amsmath}
\usepackage{amsthm}
\usepackage{amsfonts}
\usepackage{xcolor}
\usepackage{slashed}
\usepackage{mathtools}
\usepackage{epsfig}
\usepackage[colorlinks=true,citecolor=magenta]{hyperref}

\def\({\left(}
\def\){\right)}

\def\ba{\begin{eqnarray}}
\def\ea{\end{eqnarray}}

\def\be#1\ee{\begin{eqnarray}#1\end{eqnarray}}

\newcommand{\nl}{\nonumber \\}
\newcommand{\matvec}[1]{\underline{#1}}
\newcommand{\mat}[1]{\underline{\underline{#1}}}

\begin{document}

\title{Dynamics of dRGT ghost-free massive gravity in spherical symmetry}

\author{Emma Albertini}
\email{emma.albertini17@imperial.ac.uk}

\author{Jan Ko$\dot{\mathrm{z}}$uszek}
\email{j.kozuszek21@imperial.ac.uk}

\author{Toby Wiseman}
\email{t.wiseman@imperial.ac.uk}

\affiliation{Theoretical Physics Group, Blackett Laboratory, Imperial College, London SW7 2AZ, United Kingdom}

%------------------------

\begin{abstract}

We focus on dRGT massive gravity in spherical symmetry in the limit of small graviton mass.
Firstly we examine the minimal model. This does not exhibit a Vainshtein mechanism in spherical symmetry, but one may still ask what happens for spherical dynamics.
We show that there are no regular time-dependent spherically symmetric solutions unless the matter has sufficiently large pressure. For matter that does not satisfy this, such as non-relativistic matter, any Cauchy slice of such a solution must necessarily have a point where the metric becomes singular.
Only a weak assumption on the asymptotics is made. 
We then consider the next-to-minimal model. This has been argued to have a good Vainshtein mechanism in spherical symmetry, and hence be phenomenologically viable, provided the relative sign of the minimal and next-to-minimal mass terms is the same, and we restrict attention to this case.
We find that regular behaviour requires the matter at the origin of symmetry to have positive pressure -- in particular a massive scalar field fails to satisfy this condition. Furthermore it restricts non-relativistic matter so that the pressure is bounded from below in terms of the density and graviton mass in a manner that is at odds with a reasonable phenomenology. 
This suggests that realistic phenomenology will either require a resolution of singularities, or will require dynamics beyond the non-generic setting of spherical symmetry.
\end{abstract}

\maketitle

%------------------------

\section{Introduction}

In recent years many modifications to General Relativity (GR) have been considered, strongly motivated by the observation of dark matter and dark energy in cosmology. Modifications of GR that affect the long distance dynamics of the theory have the potential to provide a new perspective on whether such exotic matter is required. Perhaps the most natural long distance modification is to add a mass to the graviton.
Current model-independent bounds on the graviton mass constrain it to be less than $\sim 10^{-22}$eV. However, the phenomenologically interesting values of the mass are those close to the Hubble scale, $\sim 10^{-32}$eV, in order to potentially explain the cosmological phenomenon of dark energy~\cite{deRham:2016nuf}.

 Attempts to add a mass term to GR have a long and complicated history. In linear theory, the unique term was identified early on by Fierz and Pauli\cite{Fierz:1939ix}. However, it was shown in works by van Dam and Veltman, as well as Zakharov (vDVZ), that this would lead to observable deviations from weak-field GR, even in the limit of infinitesimally small graviton mass\cite{vanDam:1970vg,Zakharov:1970cC}. In particular, based on Solar System tests, this seemed to already exclude the possibility of a massive graviton.
 
 Vainshtein\cite{Vainshtein:1972sx} showed that this was not necessarily the case. He noted that if one takes a zero-mass limit while keeping the matter sources fixed, the metric perturbation diverges, and thus exits the regime of validity of linear theory. To properly compute the effects of massive gravity, a non-linear extension of the Fierz-Pauli mass term must thus be specified. One may then hope than in such a non-linear theory, appropriate screening mechanisms could arise that ensure the recovery of GR-like behaviour on local scales. Such a screening is now referred to as the Vainshtein mechanism (see \cite{Babichev:2013usa} for an overview).

For many years non-linear completions of the Pauli-Fierz linear theory were plagued by ghost degrees of freedom~\cite{Boulware:1973my, Aragone:1979bm, Creminelli:2005qk}. This was resolved in the work of de Rham, Gabadadze and Tolley~\cite{deRham:2010kj, deRham:2011qq}, who realized that given an additional metric structure, which we term the reference metric, there is a unique ghost-free massive extension of GR, parameterized by 3 mass terms, and a cosmological constant. The natural choice for the reference metric is to take Minkowski spacetime as the notion of a massive spin-2 field requires Poincar\'e invariance. Then by appropriately tuning the cosmological constant (as for GR), one ensures that Minkowski is a vacuum solution. This theory has been shown to propagate 5 massive graviton modes~\cite{deRham:2011qq,deRham:2011rn,Hassan:2011ea,Hassan:2011hr,Hassan:2012qv,Hinterbichler:2012cn,Kluson:2011aq,
Kluson:2011qe,Kluson:2011rt,Golovnev:2011aa,Comelli:2012vz,Kluson:2012gz,Deffayet:2012nr,Deffayet:2012zc,Kluson:2012cqq,Comelli:2013txa,Deffayet:2015rva}.

The phenomenology of this theory has been extensively studied, as reviewed in~\cite{deRham:2014zqa}. As we will discuss further below, it is claimed that one mass term leads to instabilities in the theory and therefore should be excluded~\cite{Berezhiani:2013dca}. This leaves two terms, the minimal mass term and next-to-minimal mass term. Taking only the former gives a theory that, under the conditions of spherical symmetry and stationarity, cannot exhibit the necessary Vainshtein mechanism to recover GR-like behaviour in the small mass limit~\cite{Koyama:2011yg}, although it may nonetheless have such a mechanism in less constrained situations, as emphasized in~\cite{Renaux-Petel:2014pja}. Taking both mass terms, it has been claimed that GR behaviour can be recovered on small scales~\cite{Sbisa:2012zk, Koyama:2011yg, Koyama:2011xz, Tasinato:2013rza, Comelli:2011wq, Berezhiani:2013dw, Berezhiani:2011mt, Gruzinov:2011mm,Deffayet:2008zz,Babichev:2009us,Babichev:2010jd,
Chkareuli:2011te,Babichev:2013usa} with a Vainshtein mechanism operating in spherical symmetry. We term this the non-minimal theory. However if the signs of these terms are not correlated, it is believed that this non-minimal theory suffers from a ghost excitation about such a non-linear configuration, and hence breaks down with strong coupling~\cite{Berezhiani:2013dw}. Thus phenomenological viability requires both the minimal and next-to-minimal mass terms with the correct signs.

Beyond static, spherically symmetric situations, relatively little is known about this gravitational theory~\cite{Deffayet:2001uk,deRham:2010tw,deRham:2011by, Kaloper:2011qc,
Belikov:2012xp,Hiramatsu:2012xj,Kimura:2011dc,Gannouji:2011qz,Babichev:2011iz,deRham:2012fw,deRham:2012fg,Padilla:2012ry,
DeFelice:2011th,Chu:2012kz,Andrews:2013qva,Koyama:2013paa,Li:2013nua,Dar:2018dra,Brax:2020ujo,Lara:2022gof,Bezares:2021dma,Shibata:2022gec,
TerHaar:2022wus,Dima:2021mdg}. Even for the minimal model, which at least in spherical symmetry is not expected to have a GR limit, one can nonetheless ask how the dynamics behave. For example, what happens under gravitational collapse of matter? For the non-minimal theory, which apparently has phenomenologically viable static spherical stellar solutions reproducing GR on small scales, one can ask whether those can be formed dynamically. Furthermore one can ask whether black holes that look Schwarzschild-like form generically during matter collapse~\cite{Deffayet:2011rh, Berens:2021tzd, Rosen:2017dvn}.

Progress on these questions was made in~\cite{deRham:2023ngf} where the first explicit dynamical formulation of the non-minimal (and minimal) theories was given, and spherical simulations of scalar field matter collapse were performed in the case of the minimal theory, although not in the small mass regime. Interestingly it was found that as the amplitude of the incoming matter pulse was increased, naked singularities developed generically on the time scales of the in-fall time. One motivation for this work is to move towards understanding this phenomenon.

We will focus here on both the minimal model, and also the phenomenologically viable next-to-minimal model. Further we will look at dynamics in spherical symmetry, taking the small mass limit, where `small' should be understood relative to the mass scales of the matter at the origin. It is precisely this small mass limit that is important for phenomenology.
Only a weak assumption about the asymptotic behaviour of the solutions is made, namely that the 
vierbein can be continuously deformed to its value for the Minkowski vacuum solution.
We will then find that for the minimal model no smooth Cauchy slice can exist unless the matter at the origin satisfies the very restrictive energy condition $3 P > \rho$ -- which non-relativistic matter does not. Thus if one were to try to collapse matter that violates this, initially supported away from the origin, some singular behaviour would arise. This raises the interesting question of whether this is the small mass analog  to what was seen in the numerical simulations~\cite{deRham:2023ngf}. Interestingly we will also find that, dropping spherical symmetry, regularity implies no Cauchy slice that is time-symmetric in unitary gauge can exist.

We then turn to the next-to-minimal model in spherical symmetry. Taking the viable signs for the mass terms, we find that no smooth Cauchy slice can exist unless the matter at the origin again obeys an energy condition. This requires that the pressure be positive for positive energy density, which is generically violated by  reasonable matter such as a massive scalar field. Furthermore it implies that for small but non-zero graviton mass, the pressure must be bounded below in terms of the energy density and the graviton mass in a manner that requires the matter is not \emph{too non-relativistic}. Given the density and pressure of the matter, this translates into a lower bound on the graviton length scale. Interestingly, for the matter in the Earth, regularity of dynamical solutions requires a graviton length scale comparable to the size of the observable universe.
Thus while some spherical solutions may exist which exhibit the Vainshtein mechanism and hence GR behaviour on small scales, the type of matter that supports them is severely restricted if we wish the solution to be non-singular and the graviton mass to be in the interesting cosmological range of scales.

 One might wonder whether the singular behaviour that necessarily occurs for matter that violates these stringent conditions could be tolerated. We will argue that this singular behaviour is likely to be associated to the effective theory becoming strongly coupled. To answer the question of whether one might permit such singularities in dynamics would require detailed knowledge of the full UV behaviour of the theory.

Our conclusion is certainly not that dRGT gravity cannot give rise to GR-like behaviour for realistic matter models and graviton masses that are interesting phenomenologically (i.e. smaller than the observable universe). Rather, it pushes us either to understand such singular geometries and their associated strong coupling, or to move away from the highly non-generic situation of spherical symmetry to explore the full and rich behaviour of this theory. In either case it is therefore likely that numerical as well as analytic techniques will be required to further understand this enigmatic theory; either to understand the singularities that form; or to solve the theory in the challenging dynamical regime beyond spherical symmetry.
With this in mind, the new results of~\cite{JKTW} for the minimal theory which give a formulation that is well-posed near the vacuum solution,  may well be important in developing numerical control of its dynamics.

The structure of this paper is as follows. Firstly in section~\ref{sec:theory} we review the details of the dRGT theory. Then we specify our assumption on the spatial asymptotics of the solution in section~\ref{sec:asym}. We begin our analysis in section~\ref{sec:minimal} by considering spherical symmetry in the minimal model, and showing regular behaviour constraints the energy of matter at the origin in the small mass limit. Then in the next section~\ref{sec:timesym} we find a similar restriction when considering a Cauchy surface which is a moment of time symmetry. We then consider the non-minimal theory, taking the mass parameters in the range that has previously been claimed to give a Vainshtein mechanism that can reproduce GR behaviour in the appropriate small mass limit. In that section~\ref{sec:nonminimal} we show that spherical symmetry again requires that dynamical solutions (or indeed static ones) must be supported by matter that obeys a restrictive energy condition as compared to GR. We conclude with a summary and discussion of what such singular behaviour implies for the massive dRGT theory in spherical symmetry.

\section{Ghost-free massive gravity}
\label{sec:theory}

The starting point for ghost-free massive gravity is the usual metric that matter couples minimally to, $g_{\mu\nu}$, together with a fiducial, or reference, metric $f_{\mu\nu}$. In principle this reference metric may be freely specified as that of any smooth geometry. However to ensure a clear interpretation of mass, it is natural to restrict it to be the metric of Minkowski spacetime. Here we will make this standard choice that the reference metric is that of Minkowski spacetime up to a diffeomorphism, and further we will require that the massive theory admits a Minkowski spacetime vacuum solution.
\\

To construct the massive ghost-free theory we introduce a symmetric vierbein, $E_{\mu\nu} = E_{(\mu\nu)}$ and then write the metric as~\cite{Hassan:2011vm},
\be
g_{\mu\nu} = (f^{-1})^{\alpha\beta} E_{\alpha\mu} E_{\beta\nu} 
\ee
where $f^{-1}$ is the inverse reference metric. This implies that $f^\mu_{~\nu} = E^\mu_{~\alpha} E^\alpha_{~\nu}$, where all indices are raised with the metric. 
The quantity
 $E^\mu_{~\nu} = g^{\mu\alpha} E_{\alpha\nu}$, is the building block of the massive gravity theory. Being a symmetric 2-tensor, $E_{\mu\nu}$ has the same number of components as the metric. We may interpret it as a vierbein, with the symmetry condition on its components defining a particular frame choice. It is worth noting that given a (Lorentzian) metric and reference metric it is not always possible to compute a (real) square root $\sqrt{f^\mu_{~\nu}}$ and hence find such a vierbein. However our perspective here is that our starting point is $E_{\mu\nu}$ and the reference metric, and then we always find a Lorentzian metric $g_{\mu\nu}$.

The dRGT theory with minimal and next-to-minimal mass terms is then given by the action,
\be
\label{eq:dRGTaction}
S[g, f,  \psi_i] = \frac{1}{16 \pi G_N} \int d^4x \sqrt{-g} \left(R[g] - m_{(1)}^2 (2 E^\alpha_{~\alpha}-6) - \frac{m_{(2)}^2}{2} \( (E^\alpha_{~\alpha})^2-E^{\alpha}_{~\beta} E^{\beta}_{~\alpha} -6\) \right) + S^{\rm (matter)}[g, \psi_i]\,,
\ee
where $\psi_i$ are the matter fields and $S^{\rm (matter)}$ is the usual matter action of GR where we assume minimal coupling to the metric. The mass terms, constructed from the vierbein, are the minimal  term with mass parameter $m_{(1)}^2$ and the next-to-minimal one with parameter $m_{(2)}^2$. We will use the notation $\kappa = 8 \pi G_N$ from now on.

We emphasize again that the reference metric $f_{\mu\nu}$ is chosen to be Minkowski spacetime up to a diffeomorphism. The theory as a whole is diffeomorphism invariant.
If we choose a particular coordinate system for the reference metric, for example taking $f_{\mu\nu} = \eta_{\mu\nu} \equiv \left( \begin{array}{cc} -1 & 0 \\ 0 & \delta_{ij} \end{array} \right)$, then this also fixes coordinates for the physical metric. One may make the diffeomorphism invariance of the theory manifest by writing,
\be
f_{\mu\nu} = \partial_\mu \Phi^\alpha \partial_\nu \Phi^\beta \eta_{\alpha\beta}
\ee
where $\Phi^\mu$ give the 4 components of the diffeomorphism $x^\mu \to \Phi^\mu(x)$. Then the mass terms above become kinetic terms for this vector of fields $\Phi^\mu$,  which are hence termed St\"uckelberg fields, and thought of as dynamical. Using this language,  the coordinate choice $f_{\mu\nu} = \eta_{\mu\nu}$ for the reference metric is known as taking unitary gauge.

We should emphasize that this is the leading order classical dRGT theory. Regarded as an effective field theory there will be higher dimension operators of both classical and quantum origin, controlled by some cut-off scale, $\Lambda_{cutoff}$~\cite{deRham:2017xox,deRham:2018qqo,deRham:2018bgz,Aydemir:2012nz,deRham:2014wfa}.
As for GR these include terms built from contractions of the Riemann tensor, and covariant derivatives of Riemann tensors. However they also include terms built from traces of powers of the building block matrix $\mathcal{K}^\mu_{~\nu} = \delta^\mu_{~\nu} - E^\mu_{~\nu}$, together with covariant derivatives and contractions with Riemann tensors. An example of such a set of terms is,
\be
S_{high\;dim\;op}= \Lambda^{4-3p}_{cutoff} \left( {M_{Pl} m^2} \right)^{p} \int d^4x \sqrt{-g} \mathrm{Tr}\left( \mat{\mathcal{K}}^{p} \right)
\ee
where $p$ is a positive integer, $\Lambda_{cutoff}$ is the cut-off scale of the effective field theory and $M_{Pl}^2 = 1/\kappa$. These terms are suppressed by increasing powers of the cut-off scale. However if the components of $ E^\mu_{~\nu} $ diverge at a singularity, then these terms may still come to dominate the leading classical action in~\eqref{eq:dRGTaction}. The phenomenon where higher dimension operators come to dominate the leading effective action is called strong coupling. What happens when a theory becomes strongly coupled depends on understanding the behaviour of these infinitely many higher dimension operators, and typically requires a full UV description of the theory to be known.

A naive determination of this quantum cut-off by computing loop corrections about Minkowski spacetime finds a rather small length scale $\Lambda_{cutoff}^{-1} \sim 10$km. However in regions where the Vainshtein mechanism operates, a solution must be highly non-linear and far from Minkowski spacetime, and it has been argued that loop calculations about such backgrounds will yield a much higher cut-off, possibly up to the Planck scale~\cite{deRham:2012ew,deRham:2013qqa}. Here we will treat only the leading classical theory given above.

The minimal model corresponds to $m_{(2)} = 0$, in which case the graviton mass is given by $m_{(1)}$. The non-minimal model with only the next-to-minimal mass term -- called the quadratic model as the mass term in the action is then quadratic in $E^\alpha_{~\beta}$ -- is given by taking $m_{(1)} = 0$, and then the graviton mass is given by $m_{(2)}$. 
In general the mass squared, $m^2$, of the graviton fluctuation about Minkowski is given by,
\be
m^2 = m_{(1)}^2 + m_{(2)}^2 \; .
\ee
There is a third mass term which can be added that is cubic in $E^\alpha_{~\beta}$, but this has been argued to generically lead to instabilities, and hence incompatibility with phenomenology, and so we will not consider it further here~\cite{Chkareuli:2011te,Berezhiani:2013dca,deRham:2014zqa}. Furthermore, the constraint structure of the theory with cubic term is more subtle than for the non-minimal model we focus on here, and discuss in detail shortly~\cite{Deffayet:2012nr,Bernard:2014bfa,Bernard:2015uic}.
In order to link to previous literature we may write,
\ba
m_{(1)}^2=m^2(1+2\alpha_2) \; , \quad
m_{(2)}^2=-2 m^2 \alpha_2 \; .
\ea
Then the theory is parameterized by the graviton mass $m$ and the coupling $\alpha_2$ following the notation of~\cite{deRham:2014zqa}. Furthermore, in terms of invariant couplings in that review, $\tilde{\alpha} = 2 \alpha_2$ and $\tilde{\beta} = 0$.
In order for the non-minimal theory to have a viable Vainshtein mechanism in spherical symmetry one requires that $m_{(1)}^2 > 0$ and $m_{(2)}^2 > 0$, so $-\frac{1}{2} < \alpha_2 < 0$, and we shall assume that this is the case from now onwards~\cite{Berezhiani:2013dca,Berezhiani:2013dw}. 

The Einstein equations that result from this non-minimal theory can be written as,
\be
\label{eq:EinsteinEq}
\mathcal{E}_{\mu\nu} \equiv G_{\mu\nu} + m_{(1)}^2 M^{(1)}_{\mu\nu} + m_{(2)}^2 M^{(2)}_{\mu\nu} - \kappa T_{\mu\nu} = 0\,,
\ee
where $T_{\mu\nu}$ is the usual conserved matter stress-energy tensor, and the mass terms are given explicitly as,
\ba
M^{(1)}_{\mu\nu}  & = & - E_{\mu\nu} + E^\alpha_{~\alpha} g_{\mu\nu} - 3 g_{\mu\nu} \\
M^{(2)}_{\mu\nu} & = & \frac{1}{2} E_{\mu\alpha} E^{\alpha}_{~~\nu} - \frac{1}{2} E^\alpha_{~\alpha} E_{\mu\nu} - \frac{1}{4} \left(
E^\alpha_{~\beta} E^\beta_{~\alpha}  - \left(E^\alpha_{~\alpha}\right)^2 \right) g_{\mu\nu} - \frac{3}{2} g_{\mu\nu}\,.
\ea
The cosmological constant components of these mass terms are tuned so that for any $m_{(1,2)}$ Minkowski spacetime is a solution, namely $g_{\mu\nu} = f_{\mu\nu}$, where again we emphasize that $f_{\mu\nu}$ is a diffeomorphism of Minkowski spacetime.

An important aspect of the theory is the structure of its constraints. Upon imposing stress-energy conservation and using the contracted Bianchi identity, the divergence of the Einstein equation implies,
\be
\label{eq:vector}
V_\mu = \nabla^\nu \left(   m_1^2 M^{(1)}_{\mu\nu} + m_2^2 M^{(2)}_{\mu\nu}  \right)=0
\ee
which we call the vector constraint.
Taking the small mass limit, $m_{(1)}^2$, $m_{(2)}^2 \to 0$ implies the mass terms in the Einstein equation~\eqref{eq:EinsteinEq} vanish and one naively recovers the Einstein equation of GR.
However this constraint remains non-trivial even in this limit, and is responsible for the vDVZ discontinuity in the linear theory. Likewise it is this equation that implies linear theory breaks down in the small mass limit, as Vainshtein observed. We emphasize that when we discuss this small mass limit here, and later in the paper, the scale which we are implicitly comparing the mass to is that set by the curvature, and hence by the matter which couples through the stress-energy tensor. Of course we are not literally interested in the zero mass limit, but rather the situation that the length scale associated to the mass is much larger than that set by the matter scales.

We term this equation a constraint as we see it involves one derivative of the expressions $M^{(1,2)}_{\mu\nu}$, which are algebraic in the vierbein. Hence it only depends on one derivative of the vierbein, and thus should be regarded as a constraint on the second order dynamics, and in particular a constraint on Cauchy data. 
It is convenient to define,
\be
\label{eq:xi}
\xi_\mu = E_{\mu\alpha} (f^{-1})^{\alpha\beta}V_\beta\,,
\ee
so that $\xi_\mu = 0$ is equivalent to the vector constraint vanishing. There is a further important equation, the scalar constraint~\cite{Deffayet:2012nr}. This arises from taking traces of the Einstein equation suitably combined with the divergence of $\xi$ as~\cite{deRham:2023ngf},
\be
\label{eq:scalar}
\Pi = \frac{1}{2} \left( m_1^2 g^{\mu\nu} + m_2^2 E^{\mu\nu} \right) \mathcal{E}_{\mu\nu} + \nabla \cdot \xi\,.
\ee
One finds that the second derivative terms in this contraction of the Einstein equation $\mathcal{E}_{\mu\nu}$ precisely cancel those in the divergence $\nabla \cdot \xi$, and hence this too is a constraint equation.
The scalar and vector constraints comprise 5 equations, and we may naively count the 5 physical degrees of freedom as being the 10 components of the metric minus these 5 constraints.
A more detailed and explicit exposition of the dynamics of this theory is given in the recent~\cite{deRham:2023ngf}.

In addition to these constraints, there are also the usual Hamiltonian and momentum constraints. As for GR, these are given by the $\mathcal{E}^t_{~t}$ and  $\mathcal{E}^t_{~i}$ components of the Einstein equations respectively. The contracted Bianchi identity then implies that these equations can have at most 1 time derivative -- the mass terms, being algebraic in the vierbein are irrelevant here. Provided the vector constraint is obeyed, then as for GR, these constraints are first class; they need only be imposed in the initial data, and then automatically hold under evolution. Again more details can be found in~\cite{deRham:2023ngf}.

Here we are interested in the dynamics of the minimal and non-minimal theories when the vierbein and reference metric share a symmetry. A good example of how the dynamics in the presence of symmetries can be very different from that of GR  is given by cosmological solutions~\cite{DAmico:2011jj}. Take coordinates $x^\mu = (t,x,y,z)$ and the reference metric to be Minkowski $f_{\mu\nu} = \eta_{\mu\nu}$ in its usual coordinates (so unitary gauge), and a FLRW solution for the metric, so,
\be
E_{\mu\nu} = \left( 
\begin{array}{cc}
- N(t) & 0 \\
& a(t) \delta_{ij}
\end{array}
\right)
\quad \implies \quad
g_{\mu\nu} = \left( 
\begin{array}{cc}
- N(t)^2 & 0 \\
& a(t)^2 \delta_{ij}
\end{array}
\right) \; .
\ee
Then the vector constraint, which is independent of the matter and only has a time component due to the symmetry, is equal to,
\be
\xi_t = - 3 \frac{\dot{a}}{a} \left( m_{(1)}^2 + \frac{1}{a} m_{(2)}^2 \right)
\ee
and immediately indicates that $\dot{a} = 0$. Hence homogeneity and isotropy imply a static solution for dRGT, independent of the matter chosen. As discussed in~\cite{deRham:2014zqa} this implies that cosmological solutions that could match phenomenology must have less symmetry, in either the metric or reference metric or both. For examples of such constructions, see e.g.~\cite{Kobayashi:2012fz, Gratia:2012wt, Chamseddine:2011bu}.

\section{Asymptotics}
\label{sec:asym}

Before we proceed, let us  discuss the asymptotics we will require for our theory. Here we will not make any symmetry assumption about the asymptotics of the solution. Further we will not assume that the metric is asymptotically flat -- indeed later when we consider the non-minimal theory, the weak field analysis that demonstrates it has a working Vainshtein mechanism also implies the asymptotics are not flat but cosmological, and hence time dependent~\cite{Berezhiani:2013dw,Berezhiani:2013dca}.

 We note that since the reference metric $f_{\mu\nu}$ is a diffeomorphism of the Minkowski metric $\eta_{\mu\nu}$, its signature must be the same as that of $\eta_{\mu\nu}$, so $(-,+,+,+)$. 
 The vacuum solution of the theory is $E_{\mu\nu} = f_{\mu\nu}$ and we emphasize that $E_{\mu\nu} = - f_{\mu\nu}$ is not a vacuum solution since the mass terms explicitly depend on the overall sign of the vierbein even though the metric does not.

We will make the asymptotic requirement that at spatial infinity the signature of the vierbein is the same as that of the vacuum solution, the Minkowski metric, so,
\be
\mathrm{Signature}\left( E_{\mu\nu} \right) = ( -,+,+,+) \; , \quad \mathrm{as} \; | x^i | \to \infty \; .
\ee
We may view this as the condition that we may continuously deform the vierbein tensor at any point at spatial infinity to its value in the Minkowski vacuum.
An implication of this is that the sign of the determinant of the vierbein must be the same as that of the reference metric, with both being negative asymptotically;
\be
\det\left( E_{\mu\nu} \right) < 0  \; , \quad \mathrm{as} \; | x^i | \to \infty \; .
\ee
Note that such a continuous deformation of the vierbein cannot change its signature without passing through a singularity. A change of signature would require that either its eigenvalues diverge and it becomes singular, or that some eigenvalue becomes zero, and hence it would be non-invertible. In this latter case the determinant $\det\left( E_{\mu\nu} \right)$ would vanish, but since $\det(g_{\mu\nu}) = \left( \det(E_{\mu\nu}) \right)^2 / \det(f_{\mu\nu})$ then $g_{\mu\nu}$ would also have vanishing determinant, and hence fail to be a metric and thus become singular. We emphasize that this could not be cured by a diffeomorphism. While a singular diffeomorphism may cure the zero in a metric's determinant (for example passing from polar coordinates to Cartesian coordinates does so at the origin), in the massive gravity theory it would necessarily make the reference metric singular.

This asymptotic condition then implies that for a solution to be regular in the interior of the spacetime, we should also have the same signature  for the vierbein $E_{\mu\nu}$ everywhere in this interior. The argument is the same at that above. Provided the vierbein, metric and reference metric are not singular in the interior, then we may think of passing along a curve from spatial infinity to some point in the interior as inducing a continuous and smooth change of the vierbein, and hence its signature is preserved.

\section{Spherically symmetric dynamics in the minimal theory}
\label{sec:minimal}

We now turn to examine spherically symmetric dynamics in the small mass limit, starting here in the minimal theory. By spherically symmetric dynamics we mean that both the vierbein and the reference metric  respect the same spherical symmetry. 
In fact we won't require the entire solution to be spherically symmetric, only that there is an origin of spherical symmetry, and in some neighbourhood the vierbein and reference metric respect this symmetry.
We may choose the usual spherical coordinates $x^\mu = (t, r, \theta, \phi)$, so that
\be
f_{\mu\nu} = \left( 
\begin{array}{cccc}
-1 & 0 & 0 & 0 \\
& 1 & 0 & 0 \\
& & r^2 & 0 \\
& & & r^2 \sin^2{\theta}
\end{array}
\right) \; .
\ee
We will think of this as being a spherically symmetric version of unitary gauge. The most general form for the reference metric would be a diffeomorphism of the above under $t \to T(t,r)$ and $r \to R(t, r)$, where then $T$ and $R$ would be St\"uckelberg fields.

With this choice the symmetric vierbein, which also must respect spherical symmetry, is given as,
\be
\label{eq:vierbeinspherical}
E_{\mu\nu} = \left( 
\begin{array}{cccc}
- \Phi & r \, V & 0 & 0 \\
& A + B r^2 & 0 & 0 \\
& &  A r^2 & 0 \\
& & &  A r^2  \sin^2{\theta}
\end{array}
\right)
\ee
where the component functions $\Phi$, $V$, $A$ and $B$ are functions of $t$ and $r$. Requiring smoothness at the origin, $r = 0$, imposes these component functions are smooth functions of $r^2$, and hence are even functions there.\footnote{One may check explicitly that this is the most general smooth behaviour by considering a transformation to Cartesian coordinates via the usual $(x,y,z) = (r \cos{\theta}, r \sin{\theta} \cos{\phi}, r \sin{\theta} \sin{\phi} )$ -- more detail on this may be found in~\cite{Wiseman:2011by,Adam:2011dn}.} Then the non-trivial metric components take the (smooth) form,
\be
g_{tt} = - \Phi^2 + V^2 r^2 \; , \quad g_{tr} = r V \left( \Phi + A + B r^2 \right) \; , \quad g_{rr} = \left( A + B r^2 \right)^2 - V^2 r^2 \; , \quad g_{\theta\theta} = \frac{g_{\phi\phi} }{ \sin^2{\theta}} = r^2 A^2 \; .
\ee
Assuming a regular solution in the interior, then our asymptotic condition at spatial infinity requires that the vierbein everywhere has signature $(-,+,+,+)$. Since the sign of $E_{\theta\theta}$ and $E_{\phi\phi}$ determine the sign of two of the eigenvalues of the vierbein we learn that they must both be positive, and hence;
\be
\mathrm{Signature}\left( E_{\mu\nu} \right) = ( -,+,+,+) \quad \implies \quad A > 0 \; .
\ee
The 2 by 2 block formed by  $E_{tt}$, $E_{tr}$ and $E_{rr}$ must then be Lorentzian.
The determinant of the vierbein (in the spherically symmetric region) is,
\be
\det(E_{\mu\nu}) = - r^4 A^2 \sin^2{\theta} \left( \left( A + r^2 B \right) \Phi + r^2 V^2 \right) \; .
\ee
In order to have the required signature this must be negative everywhere, including near the origin. Since $A > 0$ and near $r=0$ we have $\det(E_{\mu\nu}) = - r^4 A^3 \Phi \sin^2{\theta}  + O(r^6)$, then this implies that at the origin,
\be
 \mathrm{Signature}\left( E_{\mu\nu} \right) = ( -,+,+,+) \quad \implies \quad   \left. A \right|_{r=0} > 0 \; \mathrm{and}  \left. \Phi \right|_{r=0} > 0 \; .
\ee
We will now focus on the dynamics at the origin and expand the vierbein functions that are smooth in $r^2$ as,
\be
\Phi(t,r) = \Phi_0(t) + O(r^2) \; , \quad V(t,r) = V_0(t) + O(r^2)\; , \quad A(t,r) = A_0(t) + O(r^2)\; , \quad B(t,r) = B_0(t) + O(r^2) 
\ee
where we emphasize that $\Phi_0(t)$ and $A_0(t)$ must be strictly positive.
The vector constraint, which we recall is independent of the matter, has time component near the origin given by,
\be
\label{eq:minvecconst}
\xi_t & = & \frac{3 m_{(1)}^2}{ A_0} \left( V_0 -  \dot{A}_0 \right) + O(r^2)  \; .
\ee
Hence we determine, $V_0 = \partial_t A_0$.
Then we may substitute this into the scalar constraint giving,
\be
\label{eq:minscalconst}
0 = \frac{2 \Pi}{3 m_{(1)}^4} = \frac{1  }{ \Phi_0 } + \frac{  3 - 4 A_0  }{ A_0} - \frac{\left( \left. \kappa T^\alpha_{~\alpha} \right|_{r=0} \right)}{3 m_{(1)}^2} 
\ee 
and crucially this is linear in $\Phi_0$. Hence we may solve for $\Phi_0$ to obtain,
\be
\label{eq:minphi}
\Phi_0 = \frac{1}{4 - \frac{3}{A_0}+ \frac{\left( \left. \kappa T^\alpha_{~\alpha} \right|_{r=0} \right)}{3 m_{(1)}^2} } \; .
\ee
Now we make the further assumption that there is matter at the origin, and that this matter is generic, in the sense that $( \left. T^\alpha_{~\alpha} \right|_{r=0} )$ is finite in the small mass limit $m_{(1)}^2 \to 0$. 
Our asymptotic condition together with regularity imply that $\Phi_0, A_0 > 0$, and hence that,
\be
\mathrm{Regularity}\; \mathrm{as}\; m_{(1)}^2 \to 0 \quad \implies \quad   
 \left. \kappa T^\alpha_{~\alpha} \right|_{r=0}  \ge 0 \; .
\ee
Thus we see that regularity imposes a condition on the trace of the stress-energy tensor at the origin. While we don't know what the small mass behaviour of $A_0$ is in the denominator in the righthand side of~\eqref{eq:minphi}, and it might even diverge as $m_{(1)} \to 0$, since the term going as $-\frac{3}{A_0}$ is negative it can be ignored here; the first term in the denominator may be dropped due to the small mass limit since the stress-energy tensor term will dominate it.

Due to the spherical symmetry, smoothness implies the stress-energy tensor takes the following form near the origin,
\be
T_{\mu\nu} = \left( 
\begin{array}{cccc}
\tau & r \, \sigma  & 0 & 0 \\
& \chi + \gamma r^2 & 0 & 0 \\
& & \chi r^2 & 0 \\
& & &  \chi r^2  \sin^2{\theta}
\end{array}
\right)
\ee
where again $\tau$, $\sigma$, $\chi$ and $\gamma$ are smooth functions in $r^2$. Hence at the origin $r=0$ a co-moving observer with 4-velocity $v^\mu = ( \frac{1}{\Phi_0} ,0 ,0,0)$ sees a density $\rho_{obs}$ and a stress which is isotropic, and hence determined by a pressure $P_{obs}$, to be,
\be
\rho_{obs} =  \left. \frac{\tau}{\Phi^2}\right|_{r=0} \; , \quad P_{obs} = \left. \frac{\chi}{A^2} \right|_{r=0} 
\ee
and so we may write this condition on the trace of the stress tensor as,
\be
\label{eq:mincondition}
 \rho_{obs} \le 3 P_{obs} 
\ee
where we emphasize that we are not assuming anything about the matter; in particular the stress tensor will generally have anisotropic stress components away from the origin. In terms of an effective equation of state parameter at the origin, $P_{obs} = w \rho_{obs}$, this implies that,
\be
w \ge \frac{1}{3}  \; .
\ee
This may be viewed as a highly restrictive energy condition. In particular it is violated by non-relativistic matter where $\rho_{obs} \gg | P_{obs} |$, so $w \simeq 0$.

We already know that the minimal theory does not exhibit a Vainshtein mechanism in spherical symmetry in the low mass limit. This result is certainly in line with this, but is a much stronger statement for non-relativistic matter. No regular dynamical solution with non-relativistic matter can be found at all if the matter is not compactly supported away from the origin.

Suppose in the small mass limit the energy restriction above in~\eqref{eq:mincondition} is violated, so $T^\alpha_{~\alpha} < 0$. Then we have concluded the solution must be singular given our asymptotic condition. We may ask what the nature of such a singularity would be. We now make some general statements in the case that spherical symmetry extends to the spatial asymptotic region. Looking at \eqref{eq:minphi} for $m \to 0$, then if $T^\alpha_{~\alpha} < 0$ at the origin we must conclude that either i) $A_0 < 0$ with $A_0 \to 0^-$ as $m \to 0$, or  ii) $\Phi_0 < 0$, or iii) both of these are true. This necessarily implies the signature at the origin has changed from that at infinity, and so somewhere in between the vierbein, and hence metric, is singular. In the asymptotic region $A > 0$ and the 2 by 2 block given by $E_{tt}$, $E_{tr}$ and $E_{rr}$ must be Lorentzian. Let us denote,
\be
\Delta \equiv \det 
\left(
\begin{array}{cc}
E_{tt} & E_{tr} \\
E_{tr} & E_{rr} 
\end{array}
\right) =  - A \Phi - r^2 \left( B  \Phi + V^2 \right) 
\ee
and hence $\Delta < 0$ asymptotically. 
Then near the origin $\Delta = \Delta_0 + O(r^2)$ with $\Delta_0 = - A_0 \Phi_0$. For the cases i) and iii) above, where $A < 0$ near the origin, then $\Delta_0$ may be positive of negative depending on the sign of $\Phi$. For the remaining case ii) above, $A > 0$ near the origin, but since $\Phi < 0$ there, then $\Delta_0 > 0$.
Thus we conclude that the singularity in the vierbein is characterized by $A$ changing sign, or $\Delta$ changing sign, or both. In the interior of the solution these functions may change sign in two ways -- either discontinuously, or continuously.

In the discontinuous case derivatives of these functions will either diverge (for example if $A \sim 1/(r - r_{sing})$ then the sign of $A$ changes with $A$ being discontinuous at $r_{sing}$) or they will not be defined but $A$ will remain bounded (for example if $A \sim  \theta(r - r_{sing}) - \frac{1}{2}$). If derivatives become ill-defined then it is not clear how the effective theory would work; the discontinuity would have to be regulated in some way, which would presumably lead to the other cases, so the continuous case or the divergent discontinuous one.
If the derivatives diverge we would expect this would lead to higher derivative operators in the effective theory becoming important, and hence the theory would be strongly coupled.

The remaining option is that one or other or both of the functions $A$ and $\Delta$ vary continuously and change sign. We may compute $E^\alpha_{~\beta}$ from which $\mathcal{K}^\alpha_{~\beta} = \delta^\alpha_{~\beta} - E^\alpha_{~\beta}$ is constructed, and is the building block for higher dimension operators in the dRGT effective theory~\cite{deRham:2014zqa, deRham:2018qqo}, together with derivatives and Riemann tensors as discussed in section~\ref{sec:theory}.
We find,
\be
\label{eq:K}
\mathcal{K}^{\mu}_{~\nu} = \left(
\begin{array}{cccc}
1 + \frac{A + r^2 B}{\Delta} & \frac{r V}{\Delta} & 0 & 0 \\
- \frac{r V}{\Delta} & 1 + \frac{\Phi}{\Delta} & 0 & 0 \\
0& 0&1 -  \frac{1}{A} & 0 \\
0& 0& 0&1 -  \frac{1}{A}
\end{array}
\right) \; .
\ee
We see that if $A$ changes sign continuously, and hence at some radius becomes zero, then components of the matrix $\mathcal{K}^\mu_{~\nu}$ diverge there.\footnote{Note that we have implicitly assumed the vierbein components remain real. One might imagine a singularity where this was not the case, but then it would be very difficult to make sense of the physical behaviour. } 
Likewise if $\Delta$ changes sign continuously the same will occur. Then we expect higher dimension operators built from this matrix and its derivatives will diverge, and hence the theory will become strongly coupled.

In fact we may show that this is necessarily the case for the set of operators built simply from traces of powers of this matrix. Using matrix notation the quantity $\mathrm{Tr}( \mat{\mathcal{K}}^{2p} ) = \mathrm{Tr}( \mat{\mathcal{K}} \cdot \mat{\mathcal{K}}\cdot \ldots \cdot \mat{\mathcal{K}} )$ takes the form,
\be
\mathrm{Tr}( \mat{\mathcal{K}}^{2p} ) = 2 \left( 1 - \frac{1}{A} \right)^{2p} + \lambda_+^{2p} +\lambda_-^{2p} \; , \quad \lambda_{\pm} = \frac{F+2\Delta\pm\sqrt{F^2+4\Delta}}{2\Delta} \; , \quad F = A+Br^2+\Phi
\ee
where $p$ is a positive integer, and $\lambda_{\pm}$ are the two roots of a quadratic. If these roots are real, then we see that $\mathrm{Tr}( \mat{\mathcal{K}}^{2p} ) \ge 2 \left( 1 - \frac{1}{A} \right)^{2p}$ and hence if $A \to 0$, then this trace diverges at least as strongly as $\mathrm{Tr}( \mat{\mathcal{K}}^{2 p} ) \sim 1/A^{2p}$. 
However there is the possibility that the roots are complex, in which case $ \lambda_+^{2p} +\lambda_-^{2p}$ is real, but not necessarily positive, and further if $\Delta \to 0$ too then one could imagine a cancelation of the $\sim 2/A^{2p}$ divergence.
While this is in principle possible for some particular $p$, it cannot happen for all such terms.\footnote{When the roots are complex we may write them as, $\lambda_{\pm} = \lambda e^{ \pm i \alpha}$ for real  $\lambda$ and $\alpha$. Then we have,
\be
\mathrm{Tr}( \mat{\mathcal{K}}^{2p} ) = 2 \left(  \left( 1 - \frac{1}{A} \right)^{2p} + \lambda^{2p}  \cos( 2 p \alpha ) \right) \; .
\ee
Now for this to be finite as $A \to 0$ will require $\Delta \to 0$ so that $\lambda$ diverges and crucially also $\cos( 2 p \alpha ) < 0$. We may certainly arrange this for some particular value of $p$. However we cannot have this be true for all values of $p$;  and the traces for those $p$ then will diverge. 
To show this suppose we choose it to be true for some $p=n$ so that $3 \pi/2 > 2 n \alpha > \pi/2$. Then the cosine $\cos( 2 p \alpha )$ must be positive for either $p = 2n$ or $p= 3n$ or both.}
In the case that $A$ does not become zero, but instead $\Delta$ changes sign and goes through zero, then either one or both of $\lambda_{\pm}$ diverge as $\Delta \to 0$ and the trace again diverges. 
If the combination $F$ is non-zero, then one of $\lambda_{\pm}$ diverges as $\sim 1/\Delta$ giving a divergent trace. If $F$ vanishes simultaneously with $\Delta$ the situation is more subtle, but again one finds that the trace must diverge, at least for some values of $p$.
\footnote{Noting that, $\lambda_+ \lambda_- = (F - 1 + \Delta)/\Delta$, we find that even if $F$ vanishes simultaneously with $\Delta$, at least one of the eigenvalues must still diverge since then $\lambda_+ \lambda_- \simeq - 1/\Delta$. If $\lambda_\pm$ are real then $ \lambda_+^{2p} +\lambda_-^{2p}$ must diverge, and hence so does the trace. If they are complex, we may write them as $\lambda_{\pm} = \lambda e^{\pm i \alpha}$ for real $\lambda$ and $\alpha$, with the magnitude $\lambda$ diverging as $| \lambda |  \simeq 1/\sqrt{| \Delta |}$. Then $ \lambda_+^{2p} +\lambda_-^{2p}  = \lambda^{2 p}  \cos( 2 p \alpha ) $ will generically diverge -- it could potentially be finite for some $p$ if $\cos( 2 p \alpha ) $ is tuned to vanish, but not for all $p$.
}

\section{Time symmetric initial data in unitary gauge for the minimal theory}
\label{sec:timesym}

We now turn to consider time symmetric initial data for the minimal theory. Here by time symmetric data we refer to data on a Cauchy surface where both the vierbein and the reference metric are time symmetric.  Time symmetric Cauchy surfaces in Minkowski spacetime are hypersurfaces with constant $t$  in the usual Minkowski coordinates up to Poincare transformations. Then without loss of generality we may choose coordinates so that the reference metric is in Minkowski coordinates, so that $f_{\mu\nu} = \eta_{\mu\nu}$, and the hypersurface of time symmetry is $t = 0$.

With this choice of coordinates the vector constraint takes the form~\cite{deRham:2023ngf},
\be
\label{eq:minvector}
\xi_\mu =  - 2  (E^{-1})^{\alpha\beta} \partial_{[\mu} E_{\alpha] \beta} = 0
\ee
and the scalar constraint can be explicitly written as,
\be
\label{eq:minscalar}
2 \Pi = A^{\alpha\beta\gamma\mu\nu\rho} \partial_{[\alpha} E_{\beta] \gamma}  \partial_{[\mu} E_{\nu]\rho} + m_1^2( 3 E^\alpha_{~\alpha} - 12 ) - \kappa\, T  = 0
\ee
where we have defined the tensor,
\ba
\label{eq:A}
A^{\alpha\beta\gamma\mu\nu\rho}
&=& \eta^{\gamma\rho} g^{\alpha[ \mu} g^{\nu ]\beta} - 2 (E^{-1})^{\rho[ \alpha} g^{\beta] [\mu} (E^{-1})^{\nu]\gamma}  + 4  (E^{-1})^{\gamma[ \alpha} g^{\beta] [\mu} (E^{-1})^{\nu]\rho} \, .
\ea
Thus $\xi^\mu = 0$ is a linear constraint on $\partial_{[\alpha} E_{\beta] \sigma}$, and the scalar constraint contains a term that depends on $\partial_{[\alpha} E_{\beta] \sigma}$ as a quadratic form. 
Then assuming a smooth vierbein, we may write,
\be
E_{\mu\nu} = \left(
\begin{array}{cc}
- \phi(\vec{x}) & 0 \\
 & h_{ij}(\vec{x})
\end{array}
\right) 
+ t \left(
\begin{array}{cc}
0 & v_i(\vec{x}) \\
 & 0 
\end{array}
\right) 
+ O(t^2)
\ee
for a scalar function $\phi(\vec{x})$, vector field $ v_i(\vec{x})$ and metric $h_{ij}(\vec{x})$ on the $t=0$ hypersurface. 
Our asymptotic condition at spatial infinity, together with regularity in the interior, implies that the vierbein should everywhere have signature $(-,+,+,+)$. Further the time symmetry picks out the negative eigenvalue to be that associated to the vector $\partial/\partial t$. Thus we see that the function $\phi(\vec{x}) > 0$ must be positive, and the metric $h_{ij}(\vec{x})$ must be Euclidean.

Let us w.l.o.g. choose the point to have spatial coordinates $x^i = 0$. Then we may further simplify the metric by choosing to spatially rotate about this point so that $ h_{ij}(\vec{0})$ is diagonal, and hence,
\be
E_{\mu\nu} = \left(
\begin{array}{cccc}
- a & 0 & 0 & 0 \\
 & b_1 & 0 & 0 \\
 &  & b_2 & 0 \\
 &  &  & b_3 \\
\end{array}
\right) 
+ x^k \left(
\begin{array}{cccc}
f_k & 0 \\
 & c_{kij} 
\end{array}
\right) 
+ t \left(
\begin{array}{cccc}
0 & u_i + f_i  \\
 & 0 
\end{array}
\right) 
+ O(t^2, x^i t, x^i x^j)
\ee
with $a$, $b_i$, $c_{kij}$, $f_i$ and $u_i$ being constants computed from $\phi(\vec{x})$, $h_{ij}(\vec{x})$, $v_i(\vec{x})$ and their derivatives at $\vec{x} = 0$. Since $\phi(\vec{x}) > 0$ and $h_{ij}$ is Euclidean, this implies that,
$a > 0$ and $b_i > 0$.
Only 12 components of $\partial_{[\alpha} E_{\beta] \sigma}$  are non-zero. These are;
\begin{itemize}
\item the components $\partial_{[k} E_{i] j}$ -- there are 9 of these due to the antisymmetry in the first indices
\item the 3 components $\partial_{[t} E_{i] t}$
\end{itemize}
and we write these in terms of a spatial covector $u_i$ and the antisymmetric part of $c_{kij}$ as,
\be
c_{[ki]j} = \partial_{[k} E_{i] j} \; , \quad u_i = \partial_{[t} E_{i] t} \; .
\ee
The vector constraint has only non-zero spatial components, $\xi^i$, at $x^\mu = 0$ and hence constitutes 3 conditions. This may be used to eliminate the 3 components $u_i$ as;
\be
u_x = \frac{a}{b_y b_z} \left( b_y \left( c_{zxz} - c_{xzz} \right) + b_z \left( c_{yxy} - c_{xyy} \right)  \right) \; .
\ee
Then having satisfied the vector constraint at our point of interest, we may arrange the 9 independent components of $c_{[ki]j}$ as a vector,\footnote{
Note that we cannot freely specify the components of this vector, as it must further satisfy the constraint,
\be
V_3 - V_5 + V_7 = 0
\ee
since $E_{\mu\nu}$ is symmetric and, e.g., $c_{zxy}=c_{zyx}$.
 } 
\be
\matvec{V} = \left( c_{xyx} -c_{yxx} , c_{xyy} - c_{yxy}, c_{xyz} - c_{yxz} , c_{xzx} - c_{zxx} , c_{xzy} - c_{zxy} , c_{xzz} - c_{zxz} , c_{yzx} - c_{zyx} , c_{yzy} - c_{zyy} , c_{yzz} - c_{zyz}\right) \nl
\ee
and then we write the scalar constraint in terms of this as,
\be
\label{eq:minscalar2}
2 \Pi = \matvec{V}^T \cdot \mat{M} \cdot \matvec{V} + m_1^2( 3 [E] - 12 ) - \kappa \, T  = 0
\ee
where $\mat{M}$ is a symmetric 9 by 9 matrix, given explicitly as,
\be
\mat{M}  = \frac{1}{b_1^2 b_2^2 b_3^2} \left(
\begin{array}{ccccccccc}
2 b_3^2 & 0 & 0 & 0 & 0 & 0 & 0 & 0 & - b_1 b_3 \\
& 2 b_3^2 & 0 & 0 & 0 & b_2 b_3 & 0 & 0 & 0 \\
&  & 0 & 0 & 0 & 0 & 0 & 0 & 0 \\
&  &  & 2 b_2^2 & 0 & 0 & 0 & b_1 b_2 & 0 \\
&  &  &  & \frac{1}{2} (b_2 + b_3)^2 & 0 &  \frac{1}{2} \left( b_1 (b_2-b_3) - b_3 (b_2 + b_3 )\right) & 0 & 0 \\
&  &  &  &  & 2 b_2^2 & 0 & 0 & 0 \\
&  &  &  &  &  & \frac{1}{2}\left(b_1 +b_3 \right)^2 & 0 & 0 \\
&  \mathrm{Symmetric} &  &  &  &  &  & 2 b_1^2 & 0 \\
&  &  &  &  &  &  &  & 2 b_1^2 \\
\end{array}
\right) \; .
\ee
Being a real symmetric matrix, this must have a complete set of eigenvectors with real eigenvalues. The characteristic polynomial, whose roots determine these real eigenvalues, is,
\be
\label{eq:charpoly}
P(\Lambda) = \Lambda F_1(\Lambda) F_2(\Lambda) F_3(\Lambda) Q(\Lambda)
\ee
where $\Lambda$ is the eigenvalue, and $F_{1,2,3}$ and $Q$ are quadratics in $\Lambda$,
\be
F_1(\Lambda) &=& b_1^4 b_2^2 b_3^2 \Lambda^2 - 2 b_1^2 (b_2^2 + b_3^2) \Lambda + 3 \nl
 Q(\Lambda) & = & b_1^3 b_2^3 b_3^3 \Lambda^2 - \frac{1}{2} b_1 b_2 b_3 \left( b_1^2 + b_2^2 + 2 b_1 b_3 + 2 b_2 b_3 + 2 b_3^2 \right) \Lambda + ( b_1 + b_2 + b_3 ) 
\ee
and where $F_2$ and $F_3$ are given by taking $F_1$ and cyclically permuting $b_1$, $b_2$ and $b_3$. We know the eigenvalues for the system must be real.  Furthermore, since $b_{1,2,3} > 0$, all these four quadratics take the form,
\be
a \Lambda^2 - b \Lambda + c = 0 \; , \quad a, b, c > 0 
\ee
and from this we observe that their roots must be positive, so $\Lambda \ge 0$. 
Since the roots of the full characteristic polynomial $P(\Lambda)$ comprise the roots of these quadratics, together with the single zero eigenvalue seen from the factor $\Lambda$ in~\eqref{eq:charpoly}, then all its roots are positive. Hence all the eigenvalues of $\mat{M}$ are positive with $\Lambda \ge 0$. 
Thus we conclude that the quadratic form,
\be
\matvec{V}^T \cdot \mat{M} \cdot \matvec{V} \ge 0
\ee
is positive. Then noting that,
\be
E^\alpha_{~\alpha}  = \frac{1}{a} + \frac{1}{b_1} + \frac{1}{b_2} + \frac{1}{b_3} > 0 
\ee
 we see the scalar constraint implies,
\be
\kappa \, T =  \matvec{V}^T \cdot \mat{M} \cdot \matvec{V} + m_1^2 ( 3 E^\alpha_{~\alpha}  - 12 )   \ge - 12 m_1^2 \, .
\ee
Now for general matter at a point we have that, $T = - \rho_{obs} + P^{(1)}_{obs} + P^{(2)}_{obs} + P^{(3)}_{obs}$ where $\rho_{obs}$ is the energy density seen by an observer who is in a frame where there is no momentum flux, and $P^{(i)}_{obs}$ are the principal pressure components in that frame.
Hence we see that as $m_1 \to 0$, then the matter density and pressures must obey,
\be
P^{(1)}_{obs} + P^{(2)}_{obs} + P^{(3)}_{obs} \ge \rho_{obs}
\ee
in order for the scalar constraint to be solvable. This is a similar constraint to that required of the matter in the spherically symmetric case, and likewise is violated by non-relativistic matter. Violation of this energy condition then requires solutions to be singular. As in the spherically symmetric case, we cannot determine precisely the nature of the singularity, but again expect it to be associated to strong coupling of the theory.

\section{Spherically symmetric dynamics in the non-minimal theory}
\label{sec:nonminimal}

We now turn to the non-minimal theory, in the regime where it is claimed to be phenomenologically viable, namely $m_{(1,2)}^2 > 0$. 
It is convenient to write,
\be
m_{(1)}^2 = m^2 \alpha \; , \quad m_{(2)}^2 = m^2 \beta
\ee
so that $m$ is the graviton mass, and the constants $\alpha, \beta > 0$ are positive and satisfy $\alpha+\beta = 1$.
We will also assume these constants are fixed as we take the small mass limit -- thus $\alpha, \beta \sim O(1)$ in that limit.

Let us now repeat the analysis of Section~\ref{sec:minimal} for the minimal theory. The time component of the vector constraint near the origin is,
\be
\label{eq:nmvec}
\xi_t & = & \frac{3 m^2}{ A_0} \left( \alpha + \frac{\beta}{A_0} \right) \left( V_0 -  \dot{A}_0 \right) + O(r^2)  
\ee
so that again it is solved by the substitution $V_0 = \partial_t A_0 $.
Now substituting  into  the scalar constraint again gives a condition that is linear in $\Phi_0$ when the terms involving the stress tensor are rewritten in terms of the physical energy density and pressure measured by a co-moving observer at the origin. As in the minimal case we may simply solve for $\Phi_0$ again to obtain,
\be
\Phi_0 = \frac{\Phi_{num}}{\Phi_{denom}} 
\ee
where the numerator and denominator are,
\be
\label{eq:nmnum} 
\Phi_{num} = \alpha^2 - \alpha \beta - \frac{1}{2} \beta^2 + \frac{3 \alpha \beta}{A_0} + \frac{3 \beta^2}{2 A_0^2} + \frac{ \beta \kappa  }{3 m^2} \rho_{obs}
\ee
and
\be
\label{eq:nmdenom} 
\Phi_{denom} = 4 \alpha^2 + 2 \alpha \beta + \frac{3}{A_0} \left(  \alpha \beta - \alpha^2 + \frac{\beta^2}{2} \right) - \frac{3 \alpha \beta}{A_0^2} -  \frac{\beta^2}{2 A_0^3} + \frac{\kappa \beta}{ m^2 A_0}  P_{obs} + \frac{\kappa \alpha }{  m^2}  P_{obs}     - \frac{\kappa \alpha }{ 3 m^2}  \rho_{obs}   \; .
\ee
Now again we require both $\Phi_0 > 0$ and $A_0 > 0$ for the solution to be regular. Let us firstly assume that $\rho_{obs} > 0$. Furthermore we will assume that both $\rho_{obs}$ and $P_{obs}$ are fixed as we take the small mass limit. Then we see that the last two terms in the numerator dominate it in this limit; again we emphasize that the positive $A_0$ may have a non-trivial scaling with the mass so that it vanishes in the limit in which case the term going as $\sim 1/A_0^2$ may dominate the last one -- it would also dominate all the previous ones too.
 Either way these two terms are positive, and hence the numerator is positive. 
 
 The terms that dominate the denominator are the last four.
 The first of this set, going as $\sim 1/A_0^3$ potentially dominates the ones involving the density and pressure if $A_0 \to 0$ in the small mass limit -- it would then dominate all the prior terms in the expression. On the other hand, if $A_0$ remains finite in the small mass limit, the matter terms dominate (unless they are tuned to very precisely cancel, a non-generic situation we shall not consider further).
Since we have assumed $\rho_{obs} > 0$, then we can immediately see that, of these four terms, the only potentially positive ones are those involving $P_{obs}$, and that requires $P_{obs} > 0$. Thus we see,
\be
\rho_{obs} > 0 \; , \quad P_{obs} \le 0  \quad \implies \quad  \mathrm{lack\;of\;regularity} \; .
\ee
This is already a restrictive energy condition; if we consider a massive scalar field $\psi$ with mass $M$ and stress tensor,
\be
T_{\mu\nu} = \partial_\mu \psi \partial_\nu \psi - \frac{1}{2} g_{\mu\nu} \left( ( \partial \psi )^2 + M^2 \psi^2 \right)
\ee
then we see that for a smooth spherically symmetric scalar field $\psi = \psi(t,r)$, the density and pressure at the origin are equal to,
\be
\rho_0 = \frac{1}{2}\left. \left( M^2 \psi^2 + \frac{\dot{\psi}^2}{\Phi^2} \right) \right|_{r=0} \; , \quad P_0 = \frac{1}{2}\left. \left( - M^2 \psi^2 +\frac{\dot{\psi}^2}{\Phi^2} \right) \right|_{r=0}
\ee
and hence the density at the origin is positive and the pressure there may quite generically be negative -- in particular any time that $\dot{\psi} = 0$, so the scalar is stationary at the origin, then provided $\psi$ does not vanish, the pressure will become negative. This is generic for massive scalar field collapse in GR. For example for dispersive initial conditions, where the scalar field in-falls, and then radiates back out to infinity, the field will generically become stationary at the origin at some moment during the evolution. Requiring the scalar at the origin to vanish at the same time as being stationary would involve a fine-tuning of the initial data. 
For this non-minimal theory the expectation is that a Vainshtein mechanism does operate in spherical symmetry~\cite{Sbisa:2012zk, Koyama:2011yg, Koyama:2011xz, Tasinato:2013rza, Comelli:2011wq, Berezhiani:2013dw, Berezhiani:2011mt, Gruzinov:2011mm,Deffayet:2008zz,Babichev:2009us,Babichev:2010jd,
Chkareuli:2011te,Babichev:2013usa}. However we see here that the theory cannot reproduce the same behaviour as GR for a scalar field collapse in the small mass limit without encountering a singularity at some point in the evolution.

An important point is that in the limit $m \to 0$ the case $P_{obs} = 0 $ (as for example for dust) is singular. This indicates that for a very small but finite graviton mass, not only must $P_{obs}$ be positive in order to have a regular behaviour, but it must be bounded from below by some positive quantity that depends on the mass. This is indeed the case as we will now show.

In order to analyse this we write,
\be
\tilde{\rho} =  \frac{\kappa  \rho_{obs} }{m^2} \; , \quad \tilde{P} =  \frac{\kappa  P_{obs} }{m^2} \; .
\ee
We have assumed that $\rho_{obs} > 0$ and now further assume that $P_{obs} > 0$ (as otherwise we are guaranteed a singular solution). As above we assume that both $ \rho_{obs}$ and $P_{obs}$ are fixed as we take the mass to be small, and hence we have  $\tilde{\rho}, \tilde{P}  \to \infty$.
Now we write,
 \be
 \Phi_{denom} = Q_{denom}\left[A_0,  \tilde{P} \right] + 4 \alpha^2 + 2 \alpha \beta + \alpha \tilde{P}     - \frac{\alpha }{ 3 } \tilde{\rho} 
 \ee
 where the function $Q_{denom}$ is given as,
 \be
 Q_{denom}\left[ A_0, \tilde{P} \right] = \frac{1}{A_0^3} \left(  \left(  3 \alpha \beta - 3 \alpha^2 + \frac{3 \beta^2}{2} + \beta \tilde{P}  \right) A_0^2 - 3 \alpha \beta A_0 -  \frac{\beta^2}{2} \right) \; .
 \ee
The unknown quantity is then $A_0$, which may be very small, and in particular, may vanish in a way that scales with the mass $m$. The expression above, $\Phi_{denom}$, which must be positive for regularity, is maximized by choosing $A_0$ to maximize the function $Q_{denom}$ as $ \tilde{P} \to \infty$. 
Extremizing $Q_{denom}\left[ A_0, \tilde{P}\right] $ yields a quadratic in $A_0$, and noting that regularity requires $A_0 > 0$,  that $\alpha, \beta > 0$, and that $ \tilde{P} \to \infty$, we must choose the positive root,
\be
A_{0,max}( \tilde{P}) &=& \frac{6 \alpha \beta + \sqrt{3} \sqrt{2  \tilde{P} \beta^3 + 6 \alpha^2 \beta^2 + 6 \alpha \beta^3 + 3 \beta^4}}{2  \tilde{P} \beta - 6 \alpha^2 + 6 \alpha \beta + 3 \beta^2} \nl
& = & \sqrt{\frac{3 \beta}{2  \tilde{P}}} \left( 
1 + \frac{\sqrt{6} \alpha}{\sqrt{\beta  \tilde{P}}} + \frac{3 \left( 6 \alpha^2 -  \beta^2 - 2 \alpha \beta  \right)}{4 \beta  \tilde{P}} + O\left( \frac{1}{ \tilde{P}^{3/2}} \right) 
\right)
\ee
in which case $Q_{denom}$ is maximized, with value,
\be
\mathrm{max}( Q_{denom} ) = Q_{denom}\left[ A_{0,max}( \tilde{P}) , \tilde{P}\right]  = \sqrt{\beta} \left( \frac{2  \tilde{P}}{3} \right)^{3/2} - 2 \alpha  \tilde{P} +  \left( 2 \alpha^2 + 2 \alpha \beta + \beta^2 \right) \sqrt{\frac{3 \tilde{P}}{2 \beta}} + O(\tilde{P}^0)
\ee
as $\tilde{P} \to \infty$.
Thus $ \Phi_{denom} $ has a maximal value of,
 \be
\mathrm{max}(  \Phi_{denom}  ) &=& \mathrm{max}( Q_{denom} )  + 4 \alpha^2 + 2 \alpha \beta + \alpha \tilde{P}     - \frac{\alpha }{ 3 } \tilde{\rho} \nl
& = &  - \frac{\alpha }{ 3 } \tilde{\rho}  + \left( \sqrt{\beta} \left( \frac{2  \tilde{P}}{3} \right)^{3/2}  - \alpha \tilde{P} +  \left( 2 \alpha^2 + 2 \alpha \beta + \beta^2 \right) \sqrt{\frac{3 \tilde{P}}{2 \beta}} + O(\tilde{P}^0) \right) \; .
 \ee 
We focus on the case that $m$ is very small, and hence the first term in the brackets, going as $\sim \tilde{P}^{3/2}$, dominates the others in those brackets.
For an equation of state such as that of thermal matter or radiation, where $P_{obs} = w \rho_{obs}$ with $w = 1/3$ or $1/4$ respectively, this term will also dominate the energy density term. However, cold non-relativistic matter has a small pressure compared to its density, $| P_{obs} | \ll \rho_{obs}$. It may be that for small but finite mass, while $\tilde{P}$ is still very large, the term $\sim \tilde{P}^{3/2}$ may be comparable to, or even dominated by, the first term going as $\sim \tilde{\rho}$. Since we require that $\mathrm{max}(  \Phi_{denom}  )  > 0$ for regularity, then for small but finite mass, this yields the condition,
\be
 \sqrt{\beta} \left( \frac{2  \tilde{P}}{3} \right)^{3/2} > \frac{\alpha }{ 3 } \tilde{\rho} \; .
\ee
For relativistic matter, such as discussed above, where we expect $1 \ll | \tilde{P} | \sim \tilde{\rho}$ this is trivially satisfied. However for non-relativistic matter, $1 \ll | \tilde{P} | \ll \tilde{\rho}$, then this is a constraint that bounds the pressure from below in terms of the mass. We may say it bounds precisely \emph{how} non-relativistic the matter can be for small graviton mass in order to have regularity. 
Note that we can retrospectively check that the numerator is indeed positive as claimed earlier. Writing,
 \be
 \Phi_{num} = W_{num}\left[A_0  \right] +  \alpha^2 - \alpha \beta - \frac{1}{2} \beta^2 + \frac{ \beta^2 \kappa  }{3 m^2} \rho_{obs}
\; , \quad
W_{num}\left[A_0 \right] = \frac{3 \alpha \beta}{A_0} + \frac{3 \beta^2}{2 A_0^2}
 \ee
 then we find $W_{num}\left[ A_{0,max}( \tilde{P}) \right]  = \beta \tilde{P} + O(\sqrt{\tilde{P}})$,
 and  since $\tilde{P} \to \infty$ then $\Phi_{num}$ is indeed positive.
 
We may re-express the above regularity condition in terms of the inverse graviton mass as,
 \be
\frac{1}{m^2} > \frac{3 \alpha^2}{8 \beta} \frac{1}{ \kappa \rho_{obs}} \left( \frac{\rho_{obs}}{ P_{obs} } \right)^3  
\ee
where this holds in the small mass limit, $1 \ll \kappa \rho_{obs}/m$.
Assuming both $\alpha, \beta \sim O(1)$, and taking the Newtonian potential $\Phi_{Newt}$ at the core of the matter to parametrically go as $\Phi_{Newt} \sim P_{obs}/\rho_{obs}$,
then this bound states that,
\be
\frac{\Phi_{Newt}^3}{m^2} \gtrsim  \frac{1}{ \kappa \rho_{obs}} 
\ee
where we think of the right hand size as being fixed by the core density, and hence this bounds how small the Newtonian potential  can be (i.e. how non-relativistic the object is) for very small graviton masses. 
Taking the core density and pressure to be those for the Earth implies a bound that can only be satisfied if the length scale $m^{-1}$ is larger than $\sim 10^{26}$m, which is approximately the size of the observable universe.
This appears to imply that for regular spherically symmetric  dynamics with matter similar to that of our Earth at the origin, the graviton length scale must be approximately the size of our universe or greater. For even lower density non-relativistic objects, the bound would be even larger. Violating this bound inevitably forces the solution to be singular, and as discussed for the minimal case, we expect this singularity of the vierbein to be reflected in higher order operators diverging and the theory becoming strongly coupled.

As a final comment we consider the case that $\beta$ is negative. We emphasize that previous work has claimed that $\alpha$ and $\beta$ are required to be positive in order to obtain good phenomenology in the small mass limit of the non-minimal model~\cite{Berezhiani:2013dw}. However it may still be interesting to consider the theory in other parameter regimes. If we look at the expressions for the numerator and denominator above in~\eqref{eq:nmnum} and~\eqref{eq:nmdenom} respectively, we see that if we take $\alpha > 0$ and now $\beta < 0$, this doesn't change the sign of the terms that dominate the numerator (since they depend on $\beta^2$). However we see that of the 4 terms that may dominate the denominator, assuming the pressure is positive, the only term that is positive is the one going as $\kappa \alpha P_{obs}/m^2$. Hence unless $3P_{obs} > \rho_{obs}$, which would rule out non-relativistic matter with positive energy density entirely, the denominator cannot be positive in the small mass limit, and a singularity must necessarily exist in the solution.

\section{Summary and discussion}

We have considered spherical symmetry -- or for the minimal theory also time symmetry -- coupled with the reasonable requirement that 
the spatial asymptotics of the solution are such that the vierbein and metric may be continuously and smoothly deformed to their values for the Minkowski vacuum solution. This is a much weaker condition than asymptotic flatness, and allows for inhomogeneous, anisotropic and time dependent asymptotic behaviour.
We have then shown that if there is matter present at the origin for spherical symmetry, or at the surface of time symmetry, then in order to avoid the solution being singular this matter must be restricted to have sufficient pressure compared to its energy density for small graviton masses. 
If this is not the case, then  the vierbein and metric necessarily become non-invertible at some point in the interior of the solution and we have argued that this will be associated with strong coupling, and hence loss of control of the theory.

In the case of the minimal theory, regularity of the solution rules out non-relativistic matter entirely. For the non-minimal theory with positive mass terms, which has been claimed to be phenomenologically viable, then we find that regularity restricts the pressure to be positive -- which is already generically violated by matter such as a massive scalar field -- and further for non-relativistic matter limits how small this positive pressure can be in terms of the graviton mass and matter energy density. 
Using the density and pressure for matter at Earth's core, we find a requirement that the graviton length scale be  approximately the size of the universe or greater. If it were any smaller, a spherically symmetric solution build from Earth's matter would necessarily be singular.
Clearly this is a very different situation to that of GR, where there is no such restriction on matter in spherical symmetry.  It suggests that the  Vainshtein mechanism cannot reproduce non-singular spherically symmetric dynamical (or static) solutions with GR behaviour unless the matter is restricted in a manner that appears unreasonable to us.

Violation of these conditions on the matter necessarily leads to solutions which must contain singularities where the signature of the vierbein changes. As we have discussed, this may occur with the vierbein components diverging, in which case we expect gradients to diverge and the theory will become strongly coupled due to higher dimension operators involving derivatives  becoming important.
On the other hand, the signature may change with the vierbein varying continuously, but then the vierbein and metric will necessarily become non-invertible, and hence singular -- this cannot be undone by a singular diffeomorphism as then the singularity is shifted into the reference metric. We have argued that for global spherical symmetry this case will also become strongly coupled since the building block for higher dimension operators, $\mathcal{K}^\mu_{~\nu}$, will have diverging components.

In order to elucidate the true behaviour of these singularities, it is likely numerical methods will be required. The recent simulations of the minimal model in~\cite{deRham:2023ngf} have found singular behaviour develops in spherically symmetric collapse under time evolution away from the small mass limit. It was claimed that these singularities were associated with strong coupling. Here we argue that if one could find regular smooth initial data for matter that violates our energy conditions above, but that is initially supported away from the origin, then once it reaches the origin it will necessarily become singular. It would be interesting to test this, and see whether the singularities that develop are of a similar nature to that seen in those simulations.

In spherical symmetry the only dynamical degree of freedom in the massive theory is the spin-0 scalar mode. An important property of General Relativity is that locally we may trivialize the metric in the sense of spacetime Riemann normal coordinates. In the massive case, we may think of the theory which governs the dynamics of the diffeomorphisms of the metric, while leaving the reference metric fixed, as describing the behaviour on sufficiently small scales. This is the theory of the `decoupling limit'. In the case that only the spin-0 mode is present, such as for spherical symmetry, this is the Galileon theory~\cite{Nicolis:2008in,deRham:2010eu,Goon:2011qf,deRham:2014zqa}. As a result, the equations that we derive from the constraints, so equations~\eqref{eq:minvecconst} and~\eqref{eq:minscalconst}, which only use local vierbein data and do not refer to its second derivatives, can be equivalently thought of in this Galileon context.\footnote{We are very grateful to Andrew Tolley for pointing out this interpretation to us.} Specifically a (dual) Galileon field, $\pi$, determines a diffeomorphism as $x^\mu \to x'^\mu = x^\mu + \eta^{\mu\nu} \partial_\nu \pi(x)$. At  the origin of spherical symmetry  we may trivialize the metric to be Minkowski acted on by such a diffeomorphism by taking $E_{\mu\nu} = \eta_{\mu\nu} + \partial_\mu \partial_\nu \pi$, which results in $g_{\mu\nu} = \Lambda_\mu^{~\alpha} \Lambda_\nu^{~\beta} \eta_{\alpha\beta}$ with $\Lambda_\mu^{~\nu} = \partial_\mu x'^{\nu}$. Explicitly we take the Galileon to be a spatially spherically symmetric function of time and radius, $\pi(t,r)$, and switching to our spherical spatial coordinates and choosing, 
\be
 \pi(t,r) = \left( \frac{t^2}{2} - C_0(t) \right) + \frac{r^2}{2} \left( A_0(t) - 1 \right) + O(r^4)
\ee
where $\Phi_0(t) = C_0''(t)$ then reproduces the vierbein in equation~\eqref{eq:vierbeinspherical} and its first derivatives at the origin. Obviously it does not reproduce the vierbein behaviour away from the origin; a general metric is not a diffeomorphism of Minkowski, but will necessarily be curved. Since our arguments utilize only vector and scalar constraint equations, which are  local to the origin, we see that our equation~\eqref{eq:minvecconst} follows automatically from the Galileon form above, and the scalar constraint in equation~\eqref{eq:minscalconst} is simply the Galileon equation of motion.
While this is a reinterpretation of our derivation in the case of spherical symmetry, it would be very interesting to understand whether thinking in terms of the decoupling limit theory might allow a more direct and fruitful analysis in less symmetric situations.

Given that we have argued violating the matter restrictions leads to singular solutions that imply strong coupling, one might wonder whether this strong coupling could be tolerated. Answering this would seemingly rely on a full understanding of the UV completion of the theory. 
However, perhaps such singular and strongly coupled behaviour cannot be resolved or removed, and we should not be surprised by this given that we have already seen there are also no  FLRW solutions allowed in the theory. Hence symmetry strongly changes the behaviour of massive gravity relative to its massless sibling. Since symmetry is the epitome of non-genericity, the resolution may well be that we should forgo symmetry, at least for one of the metric or reference metric, and focus on more generic behaviour in massive gravity. It is possible that  the Vainshtein mechanism may act to recover GR-like behaviour in the small mass limit without the strong restrictions on the matter content we have seen here. Certainly testing this presents a theoretical challenge which likely will require numerical simulation. In this vein it is encouraging that very recent progress has been made towards a well-posed Cauchy evolution formulation in the absence of any symmetry~\cite{JKTW}.

\subsection*{Acknowledgments}
We are very grateful to Andrew Tolley and Claudia de Rham for many valuable discussions and for comments on this paper.
This work is supported by STFC Consolidated Grant ST/T000791/1. EA and JK are funded by STFC studentships.

\clearpage

\addcontentsline{toc}{section}{Bibliography}
\bibliography{refs}

\end{document}